\documentclass[aps,showpacs,manuscript,12pt]{revtex4}
\usepackage{amssymb}
\usepackage{amsmath}
\usepackage{graphicx}

\begin{document}
\title{\textbf{Completeness of the Lattice-Boltzmann IKT approach for
classical incompressible fluids$^{\S }$ }}
\author{E. Fonda$^{a}, $ M. Tessarotto$^{a,b}$, P. Nicolini$^{a,b}$ and M. Ellero$^{c}$}
\affiliation{\ $^{a}$Department
of Mathematics and Informatics, University of Trieste, Italy\\
$^{b}$Consortium for Magneto-fluid-dynamics\thanks{Web site:
http://cmfd.univ.trieste.it}, University of Trieste, Italy\\
$^{c}$Technical University of Munich, Germany}
\begin{abstract}
Despite the abundant literature on the subject appeared in the
last few years, the lattice Boltzmann method (LBM) is probably the
one for which a complete understanding is not yet available. As an
example, an unsolved theoretical issue is related to the
construction of a discrete kinetic theory which yields
\textit{exactly} the fluid equations, i.e., is non-asymptotic
(here denoted as \textit{LB inverse kinetic theory}). The purpose
of this paper aims at investigating discrete inverse kinetic
theories (IKT) for incompressible fluids. We intend to show that
the discrete IKT can be defined in such a way to satisfy, in
particular, the requirement of \emph{completeness}, i.e., {\it
all} fluid fields are expressed as moments of the kinetic
distribution function and {\it all} hydrodynamic equations can be
identified with suitable moment equations of an appropriate
inverse kinetic equation IKE.
\end{abstract}
\date{\today }
\maketitle



\section{Introduction}

Basic issues concerning the foundations classical hydrodynamics
still remain unanswered. A remarkable aspect is related the
construction of inverse kinetic theories (IKT) for hydrodynamic
equations in which the fluid fields are identified with suitable
moments of an appropriate kinetic probability distribution. \ The
topic has been the subject of theoretical investigations
both regarding the incompressible Navier-Stokes (NS) equations (INSE) \cite%
{Ellero2000,Ellero2004,Ellero2005,Tessarotto2006,Tessarotto2007}
and the quantum hydrodynamic equations associated to the
Schr\"{o}dinger equation\ \cite{Piero}.\ The importance of the
IKT-approach for classical hydrodynamics goes beyond the academic
interest. In fact, INSE represent a mixture of hyperbolic and
elliptic pde's, which are extremely hard to study both
analytically and numerically. As such, their investigation
represents a challenge both for mathematical analysis and for
computational fluid dynamics. The discovery of IKT
\cite{Ellero2000} provides, however, a new starting point for the
theoretical and numerical investigation of INSE. In fact, an
inverse kinetic theory yields, by definition, \emph{an exact
solver for the fluid equations}: all the fluid fields, including
the fluid pressure $p(\mathbf{r},t),$ are uniquely prescribed in
terms of suitable momenta of the kinetic distribution function,
solution of the kinetic equation. In the case of INSE this
permits, in principle, to determine the evolution of the fluid
fields without solving explicitly the Navier-Stokes equation, nor
the Poisson equations for the fluid pressure
\cite{Tessarotto2007}. Previous IKT approaches \cite%
{Ellero2004,Ellero2005,Tessarotto2006,Piero} have been based on
continuous phase-space models. However, the interesting question
arises whether similar concepts can be adopted also to the
development of discrete inverse kinetic theories based on the
lattice Boltzmann (LB) theory. The goal of this investigation is
to propose a novel LB theory for INSE, based
on the development of an IKT with discrete velocities, here denoted as \emph{%
lattice Boltzmann inverse kinetic theory (LB-IKT).} In this paper
we intend to analyze the theoretical foundations and basic
properties of the new approach useful to display its relationship
with previous CFD and lattice Boltzmann methods (LBM) for
incompressible isothermal fluids. In particular, we wish to prove
that it delivers an inverse kinetic theory, i.e., that it realizes
an exact Navier-Stokes and Poisson solver. The motivations of this
work are related to some of the basic features of customary LB
theory representing, at the same time, assets and weaknesses. One
of the main reasons of the popularity of the LB approach lays in
its simplicity and in the fact that it provides an approximate
Poisson solver, i.e., it permits to advance in time the fluid
fields without explicitly solving numerically the Poisson equation
for the fluid pressure. However customary LB approaches can yield,
at most, only asymptotic approximations for the fluid fields. This
is because of two different reasons. The first one is the
difficulty in the precise definition of the kinetic boundary
conditions in customary LBM's, since sufficiently close to the
boundary the form of the distribution function prescribed by the
boundary conditions is not generally consistent with hydrodynamic
equations. The second reason is that the kinetic description
adopted implies either the introduction of weak compressibility
\cite{McNamara1988,Higueras1989,Succi,Benzi1992,ChenpChen-1991,Chen1992}
or temperature \cite{Ansumali2002} effects of the fluid or some
sort of state equation for the fluid pressure \cite{Shi2006}.
These assumptions, although physically plausible, appear
unacceptable from the mathematical viewpoint since they represent
a breaking of the exact fluid equations. A fundamental issue is,
therefore, related to the construction of more accurate, or
higher-order, LBM's{\small ,} applicable for arbitrary values of
the relevant physical (and asymptotic) parameters. However, the
route which should permit to determine them is still uncertain,
since the very existence of an underlying exact (and
non-asymptotic) discrete kinetic
theory, analogous to the continuous inverse kinetic theory \cite%
{Ellero2004,Ellero2005}, is not yet known. According to some authors \cite%
{Shan1998,Ansumali12002,Chikatamarla2006} this should be linked to
the discretization of the Boltzmann equation, or to the possible
introduction of weakly compressible and thermal flow models.
However, the first approach is not only extremely hard to
implement \cite{Bardow}, since it is based on the adoption of
higher-order Gauss-Hermite quadratures (linked to the
discretization of the Boltzmann equation), but its truncations
yield at most asymptotic theories. Other approaches, which are
based on 'ad hoc' modifications of the fluid equations (for
example, introducing compressibility and/or temperature effects
\cite{Ansumali2005}), by definition cannot provide exact
Navier-Stokes solvers. The aim of this work is the development of
an inverse kinetic theory for the incompressible Navier-Stokes
equations (INSE) which, besides realizing an exact Navier-Stokes
(and Poisson) solver, overcomes {\tiny \ }some of the limitations
of previous LBM's. Unlike Refs. \cite{Ellero2004,Ellero2005},
where a continuous IKT was considered, here we construct a
discrete theory based on the LB velocity-space discretization. \
In such a type of approach, the kinetic description is realized by
a finite number of discrete distribution functions
$f_{i}(\mathbf{r},t)$, for $i=0,k,$ each associated to a
prescribed discrete constant velocity $\mathbf{a}_{i}$ and defined
everywhere in the existence domain of the fluid fields (the open
set $\Omega \times I$ )$.$ The configuration space $\Omega $ is a
bounded subset of the
Euclidean space $\mathbf{R}^{3}$and the time interval $I$ is a subset of $%
\mathbf{R}.$ \ The kinetic theory is obtained as\emph{\ }in \cite%
{Ellero2004,Ellero2005} by introducing an \emph{inverse kinetic
equation (LB-IKE)} which advances in time the distribution
function and by properly defining a correspondence principle,
relating a set of velocity momenta with the relevant fluid fields.

\section{2 - LB inverse kinetic theory (LB-IKT)}

There are several important motivations for seeking an exact
solver based on LBM. The lack of a theory of this type represents
in fact a weak point of LB theory. Besides being a still unsolved
theoretical issue, the problem is relevant in order to determine
the exact relationship between the LBM's and traditional CFD
schemes based on the direct discretization of the
Navier--Stokes equations. Following ideas recently developed \cite%
{Ellero2004,Ellero2005,Tessarotto2006,Piero}, we show that such a
theory can be formulated by means of an inverse kinetic theory
(IKT)
with discrete velocities. By definition such an IKT should yield \emph{%
exactly} the complete set of fluid equations and which, contrary
to customary kinetic approaches in CFD (in particular LB methods),
should not depend on asymptotic parameters. This implies that the
inverse kinetic theory must also satisfy an \emph{exact closure
condition}. As a further condition, we require that the fluid
equations are fulfilled independently of the initial conditions
for the kinetic distribution function (to be properly set) and
should hold for arbitrary fluid fields. The latter requirement is
necessary since we must expect that the validity of the inverse
kinetic theory should not be limited to a subset of possible fluid
motions nor depend on special assumptions, like a prescribed range
of Reynolds numbers. In principle a phase-space theory, yielding
an inverse kinetic theory, may be conveniently set in terms of a
quasi-probability, denoted as kinetic distribution function,
$f(\mathbf{x},t).$ A particular case of interest (investigated in
Refs.\cite{Ellero2004,Ellero2005}) refers to the case in which
$f(\mathbf{x},t)$ can actually be identified with a phase-space
probability density. In the sequel we address both cases, showing
that, to a certain extent, in both cases the formulation of a
generic IKT can actually be treated in a similar fashion. \ This
requires the introduction of an appropriate set of
\emph{constitutive assumptions} (or axioms). These concern in
particular the definitions of the kinetic equation - denoted as
\emph{inverse kinetic equation (IKE)} - which advances in time
$f(\mathbf{x},t)$ and of the velocity momenta to be identified
with the relevant fluid fields (\emph{correspondence principle}).
However,
further assumptions, such as those involving the regularity conditions for $%
f(\mathbf{x},t)$ and the prescription of its initial and boundary
conditions must clearly be added. The concept [of IKT] can be
easily extended to the case in which the kinetic distribution
function takes on only discrete values in velocity space. In the
sequel we consider for definiteness the case of the so-called
\emph{LB discretization}, whereby - for each $\left(
\mathbf{r},t\right) \in $ $\Omega \times I$ \ - the kinetic
distribution function is discrete, and in particular admits a
finite set of discrete values $f_{i}(\mathbf{r},t)\in \mathbf{R},$
for $i=0,k,$ each one corresponding to a prescribed constant
discrete velocity $\mathbf{a}_{i}\in \mathbf{R}^{3}$ for $i=0,k$.
Let us now introduce the constitutive assumptions (\emph{axioms})
set for the construction of a LB-IKT for INSE, whose form is
suggested by the analogous continuous inverse kinetic theory
\cite{Ellero2004,Ellero2005}. The axioms, define the "generic"
form of the discrete kinetic equation, its functional setting, the
momenta of the kinetic distribution function and their initial and
boundary conditions, are the following ones:

1) \ \emph{Axiom I} \emph{-} \emph{LB--IKE and functional setting.
}Let us require that the extended fluid fields $\left\{
\mathbf{V,}p_{1}\right\} $ are strong solutions of INSE, with
suitable initial and boundary conditions and that the pseudo
pressure $p_{o}(t)$ is an arbitrary, suitably smooth, real
function. In particular we impose that the fluid fields and the
volume force belong to the \emph{minimal functional setting}:

\begin{eqnarray}
&&p_{1},\Phi \epsilon C^{(2,1)}(\Omega \times I),  \nonumber \\
&&\mathbf{V}\epsilon C^{(3,1)}(\Omega \times I),  \label{Eq.6a} \\
&&\mathbf{f}_{1}\epsilon C^{(1,0)}(\Omega \times I).  \nonumber
\end{eqnarray}

We assume that in the set $\Omega \times I$ the following equation
\begin{equation}
L_{D(i)}f_{i}=\Omega _{i}(f_{i})+S_{i}  \label{Eq.7}
\end{equation}%
[\emph{LB inverse kinetic equation (LB-IKE)}]\emph{\ }is satisfied
identically by the discrete kinetic distributions $f_{i}(\mathbf{r},t)$ for $%
i=0,k.$ Here $\Omega _{i}(f_{i})$ and $L_{D(i)}$ are respectively
the BGK and the differential streaming and operators, while
$S_{i}$ is a source term to be defined.\ We require that KB-IKE is
defined in the set $\Omega \times
I,$ so that $\Omega _{i}(f_{i})$ and $S_{i}$ are at least that $%
C^{(1)}(\Omega \times I)\ $and continuous in $\overline{\Omega
}\times I.$ Moreover $\Omega _{i}(f_{i})$, to be identified as
usual with the BGK operator,\ is considered for generality and
will be useful for comparisons with customary LB approaches. \ We
remark that the choice of the equilibrium kinetic distribution
$f_{i}^{eq}$ in the BGK operator remains completely
arbitrary. We assume furthermore that in terms of $f_{i}$ the fluid fields $%
\left\{ \mathbf{V},p_{1}\right\} $ are determined by means of
functionals of the form $M_{X_{j}}\left[ f_{i}\right]
=\sum\limits_{i=0,8}X_{j}f_{i}$
(denoted as \emph{discrete velocity momenta})$.$ For $X=X_{1},X_{2}$ (with $%
X_{1}=c^{2},X_{2}=\frac{3}{\rho _{o}}\mathbf{a}_{i}$) these are
related to the fluid fields by means of the equations
(\emph{correspondence principle})
defined by the equations $p_{1}(\mathbf{r},t)-\Phi (\mathbf{r}%
)=c^{2}\sum\limits_{i=0,8}f_{i}=c^{2}\sum\limits_{i=0,8}f_{i}^{eq},$ and $%
\mathbf{V}(\mathbf{\mathbf{r},}t)\mathbf{=}\frac{3}{\rho _{o}}%
\sum\limits_{i=1,8}\mathbf{a}_{i}f_{i}=\frac{3}{\rho _{o}}\sum\limits_{i=1,8}%
\mathbf{a}_{i}f_{i}^{eq}$, where $c=\min \left\{ \left\vert \mathbf{a}%
_{i}\right\vert ,\text{ }i=1,8\right\} $ is the test particle
velocity and, without loss of generality, $f_{i}^{eq}$\ \ can be
identified with a polynomial expession, with the kinetic pressure
$p_{1}$\ replacing the fluid pressure $p$\ adopted previously
\cite{Xiaoyi1996}. These equations are assumed to hold identically
in the set $\overline{\Omega }\times I$ and by assumption, $f_{i}$
and $f_{i}^{eq}$ belong to the same functional class of real
functions defined so that the extended fluid fields belong to the
minimal functional setting (\ref{Eq.6a}). Moreover, without loss
of generality, we consider the D2Q9 LB discretization.

2) \emph{Axiom II - Kinetic initial and boundary conditions. }The
discrete kinetic distribution function satisfies, for $i=0,k$ and
for all $\mathbf{r}$
belonging to the closure $\overline{\Omega }$, the initial conditions $f_{i}(%
\mathbf{r},t_{o})=f_{oi}(\mathbf{r,}t_{o}),$ where
$f_{oi}(\mathbf{r,}t_{o})$ (for $i=0,k$) is a initial distribution
function defined in such a way to
satisfy in the same set the initial conditions for the fluid fields$p_{1o}(%
\mathbf{r})\equiv P_{o}(t_{o})+p_{o}(\mathbf{r})-\Phi (\mathbf{r}%
)=c^{2}\sum\limits_{i=0,8}f_{oi}(\mathbf{r})$ and $\mathbf{V}_{o}\mathbf{(}%
\mathbf{\mathbf{r}})=\frac{3}{\rho _{o}}\sum\limits_{i=1,8}\mathbf{a}%
_{i}f_{oi}(\mathbf{r})$ To define the analogous kinetic boundary
conditions on $\delta \Omega ,$ let us assume that $\delta \Omega
$ is a smooth, possibly moving, surface. Let us introduce the
velocity of the point of the boundary determined by the position
vector $\mathbf{r}_{w}\in \delta \Omega
, $ defined by $\mathbf{V}_{w}(\mathbf{r}_{w}(t),t)=\frac{d}{dt}\mathbf{r}%
_{w}(t)$ and denote by $\mathbf{n}(\mathbf{r}_{w},t)$ the outward
normal
unit vector, orthogonal to the boundary $\delta \Omega $ at the point $%
\mathbf{r}_{w}.$ Let us denote by $f_{i}^{(+)}(\mathbf{r}_{w},t)$ and $%
f_{i}^{(-)}(\mathbf{r}_{w},t)$ the kinetic distributions which
carry the
discrete velocities $\mathbf{a}_{i}$ for which there results respectively $%
\left( \mathbf{a}_{i}-\mathbf{V}_{w}\right) \cdot \mathbf{n}(\mathbf{r}%
_{w},t)>0$ (outgoing-velocity distributions) and $\left( \mathbf{a}_{i}-%
\mathbf{V}_{w}\right) \cdot \mathbf{n}(\mathbf{r}_{w},t)\leq 0$
(incoming-velocity distributions) and which are identically zero
otherwise.
We assume for definiteness that both sets, for which $\left\vert \mathbf{a}%
_{i}\right\vert >0,$ are non empty (which requires that the
parameter $c$ be suitably defined so that $c>\left\vert
\mathbf{V}_{w}\right\vert $). The boundary conditions are obtained
by suitably prescribing the incoming kinetic distribution
$f_{i}^{(-)}(\mathbf{r}_{w},t),$ i.e., imposing (for
all $\left( \mathbf{r}_{w},t\right) \in \delta \Omega \times I$) $%
f_{i}^{(-)}(\mathbf{r}_{w},t)=f_{oi}^{(-)}(\mathbf{r}_{w},t).$ Here $%
f_{oi}^{(-)}(\mathbf{r}_{w},t)$ are suitable functions, to be
assumed
non-vanishing and defined only for incoming discrete velocities for which $%
\left( \mathbf{a}_{i}-\mathbf{V}_{w}\right) \cdot \mathbf{n}(\mathbf{r}%
_{w},t)\leq 0$.

3) \emph{Axiom III -} \emph{Moment equations. }If
$f_{i}(\mathbf{r},t),$ for $i=0,k,$ are arbitrary solutions of
LB-IKE [Eq.(\ref{Eq.7})] which satisfy Axioms I and II validity of
Axioms I and II, we assume that the moment
equations of the same LB-IKE, evaluated in terms of the moment operators $%
M_{X_{j}}\left[ \cdot \right] =\sum\limits_{i=0,8}X_{j}\cdot ,$
with $j=1,2,$ coincide identically with INSE, namely that there
results identically [for all $\left( \mathbf{r},t\right) \in
\Omega \times I$] $M_{X_{1}}\left[
L_{i}f_{i}-\Omega _{i}(f_{i})-S_{i}\right] =\nabla \cdot \mathbf{V}=0$ and $%
M_{X_{2}}\left[ L_{i}f_{i}-\Omega _{i}(f_{i})-S_{i}\right] =N\mathbf{V}=%
\mathbf{0.}$

4) \emph{Axiom IV - Source term. }The source term is required to
depend on a finite number of momenta of the distribution function.
It is assumed that
these include, at most, the extended fluid fields $\left\{ \mathbf{V,}%
p_{1}\right\} $ and the kinetic tensor pressure $\underline{\underline{\Pi }}%
=3\sum_{i=0}^{8}f_{i}\mathbf{a}_{i}\mathbf{a}_{i}-\rho
_{o}\mathbf{VV}.$ Furthermore,\ we also normally require\emph{\
}(except for the LB-IKT described in Appendix B)\emph{\ }that
$S_{i}(\mathbf{r},t)$ results
independent of $f_{i}^{eq}(\mathbf{r,}t),$ $f_{oi}(\mathbf{r})$ and $f_{wi}(%
\mathbf{r}_{w},t)$ (for $i=0,k$).

Although, the implications will made clear in the following
sections, it is manifest that these axioms do not specify uniquely
the form (and functional
class) of the equilibrium kinetic distribution function $f_{i}^{eq}(\mathbf{%
r,}t),$ nor of the initial and boundary kinetic distribution
functions
suitably defined. Thus, both $f_{i}^{eq}(\mathbf{r,}t),f_{oi}(\mathbf{r,}%
t_{o})$ and the related distribution they still remain in principle \emph{%
completely arbitrary}. \ Nevertheless, by construction, the
initial and (Dirichlet) boundary conditions for the fluid fields
are satisfied identically. \ In the sequel we show that these
axioms define a (non-empty) family of parameter-dependent
LB-IKT's, depending on two constant free parameters $\nu _{c},c>0$
and one arbitrary real function $P_{o}(t).$ The examples
considered are reported respectively in the following Sec. 5,6 and
in the Appendix B.

\section{3 - A possible realization: the integral LB-IKT}

We now show that, for arbitrary choices of the distributions $f_{i}(\mathbf{%
r,}t)$ and $f_{i}^{eq}(\mathbf{r,}t)$ which fulfill axioms I-IV$,$
an explicit (and non-unique) realization of the LB-IKT can
actually be obtained. We prove, in particular, that a possible
realization of the discrete inverse kinetic theory, to be denoted
as \emph{integral LB-IKT, }is provided by the source term
\cite{Tessarotto2007b}

\begin{eqnarray}
&&\left. S_{i}=\right.   \label{Eq.14} \\
&\equiv &\frac{w_{i}}{c^{2}}\left[ \frac{\partial p_{1}}{\partial t}-\mathbf{%
a}_{i}\cdot \left( \mathbf{f}_{1}\mathbf{-}\mu \mathbf{\nabla }^{2}\mathbf{V}%
-\nabla \cdot \underline{\underline{\Pi }}+\nabla p\right) \right]
\equiv \widetilde{S}_{i},  \nonumber
\end{eqnarray}%
where $\frac{w_{i}}{c^{2}}\frac{\partial p_{1}}{\partial t}$ is
denoted as first pressure term. Then the following
theorem hols.\\

{\bf Theorem - Integral LB-IKT}

\emph{In validity of axioms I-IV the following statements hold.
For an arbitrary particular solution }$f_{i}$\emph{\ and for
arbitrary extended fluid fields}$:$\emph{\ } \emph{A) if }$\
f_{i}$\ \emph{\ is a solution of LB-IKE [Eq.(\ref{Eq.7})] the
moment equations coincide identically with INSE in the set
}$\Omega \times I;$ \emph{B) the initial conditions and the
(Dirichlet) boundary conditions for the fluid fields are satisfied
identically;} \emph{C) in validity of axiom IV the source term
}$\widetilde{S}_{i}$\emph{\ is non-uniquely defined by
Eq.(\ref{Eq.14}).}

\textbf{Proof}

\emph{A)} We notice that by definition there results identically

\begin{equation}
\sum_{i=0}^{8}\widetilde{S}_{i}=\frac{1}{c^{2}}\frac{\partial p_{1}}{%
\partial t}  \label{Eq.19}
\end{equation}%
$\mathbf{\qquad }$%
\begin{eqnarray}
&&\left. \sum_{i=0}^{8}\mathbf{a}_{i}\widetilde{S}_{i}=\right.
\label{Eq.20} \\
&&\left. =-\frac{1}{3}\left[ \mathbf{f-}\mu \mathbf{\nabla }^{2}\mathbf{V-}%
\nabla \cdot \underline{\underline{\Pi }}+\nabla p\right] \right.
\nonumber
\end{eqnarray}%
\ On the other hand, by construction (Axiom I) $f_{i}$ ($i=1,k$)
is defined
so that there results identically $\sum_{i=0}^{8}\Omega _{i}=0$ and $%
\sum_{i=0}^{8}\mathbf{a}_{i}\Omega _{i}=\mathbf{0}.$ Hence the momenta $%
M_{X_{1}},M_{X_{2}}$ of LB-IKE deliver respectively
\begin{equation}
\nabla \cdot \sum\limits_{i=1,8}\mathbf{a}_{i}f_{i}=0  \label{21}
\end{equation}%
\begin{equation}
3\frac{\partial }{\partial t}\sum\limits_{i=1,8}\mathbf{a}_{i}f_{i}+\rho _{o}%
\mathbf{V\cdot \nabla V+\nabla }p_{1}+\mathbf{f-}\mu \mathbf{\nabla }^{2}%
\mathbf{V}=\mathbf{0}  \label{22}
\end{equation}%
where the fluid fields $\mathbf{V,}p_{1}$ are defined by
appropriate moments \cite{Xiaoyi1996}. Hence Eqs.(\ref{21}) and
(\ref{22}) coincide respectively with the isochoricity and
Navier-Stokes equations. As a consequence, $f_{i}$
is a particular solution of LB-IKE iff the fluid fields $\left\{ \mathbf{V,}%
p_{1}\right\} $ are strong solutions of INSE.

\emph{B)} Initial and boundary conditions for the fluid fields are
satisfied identically by construction thanks to Axiom II.

\emph{C) }However, even prescribing $\nu _{c},c>0$ and the real function $%
P_{o}(t)$, the functional form of the equation cannot be unique
The non
uniqueness of the functional form of the source term $\widetilde{S}_{i}(%
\mathbf{r},t)$ is assumed to be independent of
$f_{i}^{eq}(\mathbf{r,}t)$
[and hence of Eq.(\ref{Eq.7})] is obvious. In fact, let us assume that $%
\widetilde{S}_{i}$ is a particular solution for the source term
which
satisfies the previous axioms I-IV. Then, it is always possible to add to $%
S_{i}$ arbitrary terms of the form $\widetilde{S}_{i}+\delta S_{i},$ with $%
\delta S_{i}\neq 0$ which depends only on the momenta indicated
above, and
gives vanishing contributions to the first two moment equations, namely $%
M_{X_{j}}\left[ \delta S_{i}\right]
=\sum\limits_{i=0,8}X_{j}\delta S_{i}=0,$ with $j=1,2$. \ To prove
the non-uniqueness of the source term $S_{i}$, it is sufficient to
notice that, for example, any term of the form $\delta
S_{i}=\left( \frac{3}{2}\frac{a_{i}^{2}}{c^{2}}-1\right)
F(\mathbf{r},t)$, with $F(\mathbf{r},t)$ an arbitrary real
function (to be assumed, thanks to Axiom IV, a linear function of
the fluid velocity), gives vanishing contributions to the momenta
$M_{X_{1}},M_{X_{2}}.$ Hence $\widetilde{S}_{i}$ is non-unique.

The implications of the theorem are straightforward. First,
manifestly, it holds also in the case in which the BGK operator
vanishes identically. This occurs letting $\nu _{c}=0$ in the
whole domain $\Omega \times I.$ Hence the
inverse kinetic equation holds independently of the specific definition of $%
\ f_{i}^{eq}(\mathbf{r,}t).$

An interesting feature of the present approach lies in the choice
of the boundary condition adopted for $f_{i}(\mathbf{r,}t),$ which
is different from that usually adopted in LBM's [see for example
\cite{Succi} for a review on the subject]. In particular, the
choice adopted is the simplest permitting to fulfill the Dirichlet
boundary conditions [imposed on the
fluid fields]. This is obtained prescribing the functional form of $f_{i}(%
\mathbf{r,}t)$ on the boundary of the fluid domain ($\delta \Omega
$), which is identified with a function $f_{oi}(\mathbf{r},t).$

Second, the functional class of $f_{i}(\mathbf{r,}t),$ $f_{i}^{eq}(\mathbf{r,%
}t)$ and of $f_{oi}(\mathbf{r},t)$ remains essentially arbitrary.
Thus, in particular, the initial and boundary conditions,
specified by the same function $f_{oi}(\mathbf{r},t),$ can be
suitably defined. As further basic consequence,
$f_{i}^{eq}(\mathbf{r,}t)$ and $f_{i}(\mathbf{r,}t)$ need not
necessarily be Galilei-invariant (in particular they may not be
invariant with respect to velocity translations), although the
fluid equations must be necessarily fully Galilei-covariant. As a
consequence it is always possible to select\
$f_{i}^{eq}(\mathbf{r,}t)$ and $f_{oi}(\mathbf{r},t)$ based on
convenience and mathematical simplicity. Thus, besides
distributions which are Galilei invariant and satisfy a principle
of maximum entropy (see for example \cite{Karlin1998,Karlin1999}),
it is always possible to identify them [i.e., $\
f_{i}^{eq}(\mathbf{r,}t),f_{oi}(\mathbf{r},t)$] with a
non-Galilean invariant polynomial distribution. We mention that
the non-uniqueness of the source term $\widetilde{S}_{i}$ can be
exploited also by imposing that $f_{i}^{eq}(\mathbf{r,}t)$ results
a particular solution of
the inverse kinetic equation Eq.(\ref{Eq.7}) and there results also $f_{oi}(%
\mathbf{r},t)=f_{i}^{eq}(\mathbf{r,}t)$.

\section{4 - Conclusions}

In this paper we have presented the theoretical foundations of a
new phase-space model for incompressible isothermal fluids, based
on a generalization of customary lattice Boltzmann
approaches.{\small \ }We have shown that many of the limitations
of traditional (asymptotic) LBM's can be overcome. \ As a main
result, we have proven that the \emph{LB-IKT}\textit{\ \ }can be
developed in such a way that it furnishes exact Navier-Stokes and
Poisson solvers, i.e., it is - in a proper sense - an inverse
kinetic theory for INSE. \ The theory exhibits several features,
in particular we have proven that the integral LB-IKT (see Sec.3):

\begin{enumerate}
\item determines uniquely the fluid pressure $p(\mathbf{r},t)$ via the
discrete kinetic distribution function without solving explicitly
(i.e., numerically) the Poisson equation for the fluid pressure.
Although analogous to traditional LBM's, this is interesting since
it is achieved without introducing compressibility and/or thermal
effects. In particular the present theory does not rely on a state
equation for the fluid pressure.

\item is \emph{complete}, namely all fluid fields are expressed as momenta
of the distribution function and all hydrodynamic equations are
identified with suitable moment equations of the LB inverse
kinetic equation.

\item allows arbitrary initial and boundary conditions for the fluid fields.

\item is \emph{self-consistent}: the kinetic theory holds for arbitrary,
suitably smooth initial conditions for the kinetic distribution
function. In other words, the initial kinetic distribution
function must remain arbitrary even if a suitable set of its
momenta are prescribed at the initial time.

\item the associated the kinetic and equilibrium distribution functions can
always be chosen to belong to the class of non-Galilei-invariant
distributions. In particular the equilibrium kinetic distribution
can always be identified with a polynomial of second degree in the
velocity.

\item is \emph{non-asymptotic}, i.e., unlike traditional LBM's it does not
depend on any small parameter, in particular \emph{it holds for
finite Mach numbers}.
\end{enumerate}

The main result of the paper is represented by the construction of
an explicit realization of the LB-IKT for the incompressible
Navier-Stokes equations. The construction of a discrete inverse
kinetic theory of this type for the incompressible Navier-Stokes
equations represents an exciting development for the phase-space
description of fluid dynamics, providing a new starting point for
theoretical and numerical investigations based on LB theory. In
our view, the route to more accurate, higher-order LBM's, here
pointed out, will be important in order to achieve substantial
improvements in the efficiency of LBM's in the near future.


\bigskip 

\section*{Acknowledgments}
Work developed (M.T.) in the framework of the MIUR (Italian
Ministry of University and Research) PRIN Research Program
``Modelli della teoria cinetica matematica nello studio dei
sistemi complessi nelle scienze applicate'' and the European COST
action P17 (M.T). The partial of the GNFM (National Group of
Mathematical Physics) of INDAM (National Institute of Advanced
Mathematics, Italy) (M.T. and P.N.) and of the Deutsche
Forschungsgemeinschaft via the project EL503/1-1 (M.E.) is
acknowledged.

\section*{Notice}
$^{\S }$ contributed paper at RGD26 (Kyoto, Japan, July 2008).
\newpage



\end{document}